# Lattice phonon modes of the spin crossover crystal [Fe(phen)$_2$(NCS)$_2$] studied by THz, IR, Raman spectroscopies and DFT calculations

Eric Collet*,[a]   Giovanni Azzolina,[a]   Tomoaki Ichii,[b]   Laurent Guerin,[a]   Roman Bertoni,[a] Alain Moréac,[a] Marco Cammarata,[a] Nathalie Daro,[c] Guillaume Chastanet,[c] Jacek Kubicki,[d] Koichiro Tanaka,[b] and Samir F. Matar[c,e]

[a]   Univ Rennes, CNRS, IPR (Institut de Physique de Rennes), UMR 6251, F-35000 Rennes, France

[b]   Department of Physics, Graduate School of Science, Kyoto University, Kitashirakawa-oiwakecho, Sakyo-ku, Kyoto 606-8502, Japan

[c]   CNRS, University of Bordeaux, ICMCB, UMR 5026, 87, avenue du Docteur-Albert-Schweitzer, 33608 Pessac, France

[d]   Adam Mickiewicz University in Poznan, Faculty of Physics, Umultowska 85, 61-614 Poznan, Poland

[e]   Lebanese German University (LGU) Sahel Alma Campus, P.O. BOX 206 Jounieh, Lebanon

email: eric.collet@univ-rennes1.fr

**Abstract:** [Fe(phen)$_2$(NCS)$_2$] is a prototype transition metal complex material, which undergoes a phase transition between low-spin (LS) and high-spin (HS) phases, induced by temperature, pressure or light. Vibrational modes play a key role for spin-state switching both in thermal and photo-induced cases, by contributing to vibrational entropy for thermal equilibrium transitions or driving the fast structural trapping of the photoinduced high spin state. Here we study the crystal phonon modes of [Fe(phen)$_2$(NCS)$_2$], by combining THz, IR, and Raman spectroscopies sensitive to modes in different frequency ranges and different symmetries. We compare the experimental results to DFT calculations performed in a periodic 3D crystal for understanding the phonon modes in the crystal, compared to molecular vibrations. Indeed, each vibrational mode of the isolated molecule combines into several modes of different symmetry and frequency in the crystal, as the unit cell contains four molecules. We focus our attention on the HS symmetric and anti-symmetric breathing modes in the crystal as well as on the N-CS stretching modes.

## 1 Introduction

Phonon modes are known for playing an important role during thermal phase transition in many materials. The softening of optical modes can lead to symmetry breaking, whereas the change in frequencies between phases is responsible for a vibrational contribution to the entropy change, which may drive phase transition. This is also true for transition metal complexes, which undergo spin-state switching from low-spin (LS) to high-spin (HS) state [1]. These so-called spin-crossover materials (SCO) have attracted widespread attention because they represent testbed molecular-based examples for understanding mechanisms involved during phase transitions at thermal equilibrium or out-of-equilibrium dynamics induced by light, which provide opportunities for future and emerging technologies. The [Fe(phen)$_2$(NCS)$_2$] material studied here (Fig. 1) is a well-known prototypical system and its spin state switching induced by light irradiation, temperature and pressure was intensively investigated by various techniques [2-16]. It undergoes a first-order phase transition around ≈180 K from LS (S=0, $t_{2g}^6 e_g^0$) to HS (S=2, $t_{2g}^4 e_g^2$) states. It also exhibits photomagnetic and photochromic properties, long-lived below ≈ 60 K, or transient above [2-4,7,8,11,17,18]. The molecular bi-stability of this compound is associated with important structural reorganizations between both spin states [4,7,9]. The less bonding character of the HS state leads to the expansion of the Fe-ligand distance from $<Fe-N>_{LS}$= 1.97 Å to $<Fe-N>_{HS}$=2.16 Å, often reported for Fe$^{II}$ SCO [19]. This structural reorganization gives rise to an energy barrier between the HS and the ground LS state. The LS to HS conversion at thermal equilibrium comes from the entropy change between LS and HS phases, due on the one hand to the higher spin multiplicity of the HS state and on the other hand to a vibrational entropy contribution, which comes from the shift of vibrational frequencies upon the spin crossover [5,20-24]. For photoinduced spin-state switching, it has been shown that the light-induced excited spin-state trapping (LIESST) process results from the ultrafast activation and damping of the molecular breathing mode, related to the elongation of the Fe-N bonds,[3,4,25-28] a process also responsible in crystals for elastically-driven cooperative transformation [2,29]. The important electron-phonon coupling is at the heart of thermally induced and photoinduced spin transition. It may also open new possibilities for driving spin conversion by vibrational excitation [30,31]. Gaining knowledge on the phonon modes in such bistable molecular crystals is therefore of interest, but vibrations were mainly discussed so far in terms of intramolecular modes. Here we combine IR, Raman and time-domain terahertz (TD-THz) spectroscopies, sensitive to modes in different frequency ranges or of different symmetries, and DFT calculations in a periodic 3D crystal, for investigating the lattice phonon modes of the [Fe(phen)$_2$(NCS)$_2$] crystalline system.

# Phonon modes of the spin crossover complex [Fe(phen)$_2$(NCS)$_2$]

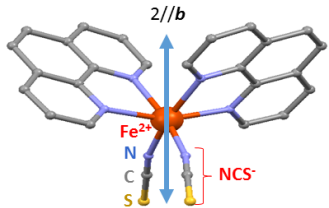

[Fe(phen)$_2$(NCS)$_2$]
molecule
point group C$_2$ (2//***b***)

| C$_2$ | E | C$_2$ | linear functions, rotations | quadratic functions |
|---|---|---|---|---|
| A | +1 | +1 | y, R$_y$ | x$^2$, y$^2$, z$^2$, xz |
| B | +1 | −1 | x, z, R$_x$, R$_z$ | xy, yz |

**Figure 1.** Structure of the [Fe(phen)$_2$(NCS)$_2$] molecule with the table of its C$_2$ point group. Vibration modes belong to A or B symmetry. For IR-active modes, the dipole moment is oriented along ***b*** for A modes (y component) or is perpendicular for B modes (*x,z* component). Hydrogen atoms are not shown.

## 2 Experimental Section

Single crystals of [Fe(phen)2(NCS)2] compound were obtained by slow diffusion as indicated by Real *et al.* [18]. The FeSO4.7H$_2$O, KNCS and 1,10-phenanthroline (phen) reactants were used as purchased. We purified the methanol by distillation over Mg under inert atmosphere prior to use. A solution at 1.10$^{-3}$ mol of Fe(NCS)$_2$ in 10 mL of methanol (from the reaction of Fe(SO$_4$)$_2$.7H$_2$O and KNCS) and a solution at 2.10$^{-3}$ mol of 1,10 phenanthroline in 10 mL of methanol were firstly prepared. The single crystals were obtained after few days by slow diffusion of the two solutions in a H-shape tube under inert atmosphere (1 or 2 mL of the two reactants separated with methanol).

We investigated low frequency IR-active modes by time-domain THz spectroscopy. We used an air plasma method for the generation of THz pulses by using an 800 nm laser pulse (Ti: sapphire regenerative amplifier, 35 fs, 1.7 mJ, 1 kHz) [32]. The measurements were performed on a single crystal of [Fe(phen)$_2$(NCS)$_2$], in the HS state and for THz polarization parallel to the crystalline axis ***a*** and ***b***. The transmitted THz radiation was measured by air-biased-coherent-detection (ABCD) method [33]. The IR absorption spectrum was collected using a commercial FTIR spectrometer (Tensor 27). For this measurement, small [Fe(phen)$_2$(NCS)$_2$] crystals were dispersed in KBr pellet. The studies by Micro-Raman Spectroscopy were carried out on a LabRAM HR800 (Horiba Scientific, Jobin-Yvon). The HR800 system is a high-resolution spectrometer fitted with a confocal microscopy system coupled to different lasers. Our experiments were performed with a 633 nm He-Ne laser as excitation source. To avoid sample heating, an optical density filter was used, the excitation laser power focused under the objective was approximately 0.1 mW. Both the THz-TDS, IR and Raman experiments were performed at room temperature in the HS phase.

For theoretical calculations of the phonon modes in the crystal, we used the VASP package [34] within density functional theory (DFT) [35]. It provides a way to get zero entropy (T = 0K) quantum mechanical energies and forces for atoms in crystals. We performed massively parallel calculations on Xeon Linux 32 processors workstations (collaboration with the University of Bordeaux computer centre MCIA: Mésocentre de Calcul Intensif en Aquitaine). On a fully geometry relaxed RT structure the frequencies were obtained by small displacements (~0.015 Å) and the resulting frequencies are written to the output. This involves to calculate the forces between atoms in the crystal and to construct the force constant matrix. Subsequently force constant matrix allows calculating the normal modes. Within VASP-DFT to Hellman-Feynman theorem is used and this method of calculating the force constant matrix by explicitly displacing atoms is called the Frozen-Phonon method. It is generally quicker and computationally cheaper than the linear response method, which utilizes density functional perturbation theory to calculate forces (unlikely to be used for our SCO crystal). Use was made of Projector augmented wave PAW method [36] built within the generalized gradient approximation (GGA) [37]. Exploiting the OUTCAR VASP-output with WxDragon software (http://www.wxdragon.de/) allows visualizing the different vibrations and record movies (supplementary information), and found the (pseudo) symmetry of the modes in the crystalline structure. Confrontation with molecular Gaussian09 [38] calculation results may allow rationalizing the differences molecule–solid; for instance by comparing the different TD-THz spectroscopy data measured with polarization along x and y axes. It also has to be underlined that we obtained, for the molecule,[3] the UV-VIS absorption spectra from the computation of excited states based on molecular Gaussian09 [38] code thanks to time dependent DFT (TD-DFT). Further details can be found in a review article [39]. The [Fe(phen)$_2$(NCS)$_2$] SCO material crystallises in the *Pbcn* space group. Because of the large size of the unit cell, which contains four molecular units (i.e. 204 atoms and 612 modes including 609 optical phonon modes), we had to exclude the 64 hydrogen atoms of the unit cell. We performed then calculations on the 140 non-hydrogen atoms (with 420 modes including 417 optical phonon modes). This is acceptable since our discussion focuses on the symmetry and the nature of low frequency breathing or bending molecular modes, as well as on the higher frequency N–CS stretching modes at about 2060 cm$^{-1}$, i.e. well below C–H stretches at about 3000 cm$^{-1}$. In our calculation, some modes are found to have negative frequency due to some instabilities (or to the lack of H atoms). In addition, we carried out calculations only at the Γ point of the Brillouin Zone (BZ), which is relevant for analysing IR and Raman data. It implies that we did not span the main directions of the primitive orthorhombic BZ, i.e. without the possibility of getting phonon density of states nor band dispersion. The drawback is that we cannot directly distinguish Raman and Infrared active modes, neither assign symmetry in the *mmm* point group of the *Pbcn* crystalline structure (A$_g$, A$_u$, B$_{1g}$, B$_{2g}$, B$_{3g}$, B$_{1u}$, B$_{2u}$, B$_{3u}$) for each frequency. However, the visualization of the atomic motions (supplemental videos) for the different modes obtained in this way allows distinguishing the symmetry of the modes as discussed hereafter.

## 3 Results and Discussion

Before discussing and interpreting IR, THz and Raman spectroscopy results with DFT calculations, a discussion about the symmetry of the [Fe(phen)$_2$(NCS)$_2$] molecular and crystalline structures of this SCO material is required.

### 3.1 Symmetry of the molecular modes

Figure 1 shows the isolated [Fe(phen)$_2$(NCS)$_2$] molecule. Its point symmetry group is C$_2$ and it is made of 51 atoms (C$_{26}$H$_{16}$FeN$_6$S$_2$). Its 153 modes include 3 translations, 3 rotations and 147 intra-molecular vibrations. The harmonic vibrations belonging to irreducible representation A are symmetric to both the identity operation *E* as well as the 180-degree rotation around the *C$_2$* axis.

# Phonon modes of the spin crossover complex [Fe(phen)$_2$(NCS)$_2$]

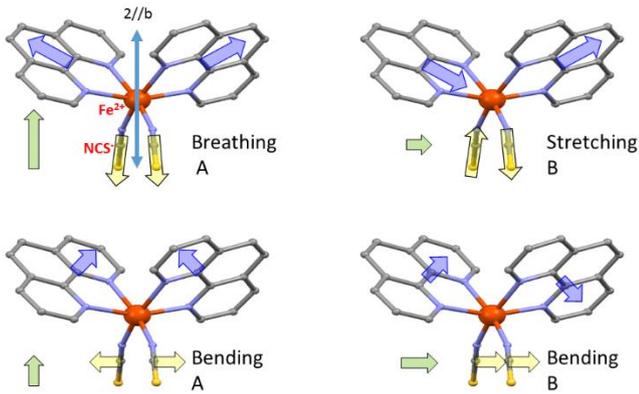

**Figure 2.** Schematic representation of low frequency modes decomposed in terms of motions of the *phen* groups (Fe–phen blue arrows) and Fe–NCS groups (yellow arrows). The variation of the dipole moment (green arrow) induced by these motions is oriented along **b** (y component) for A modes and perpendicular to **b** (x,z component) for B modes.

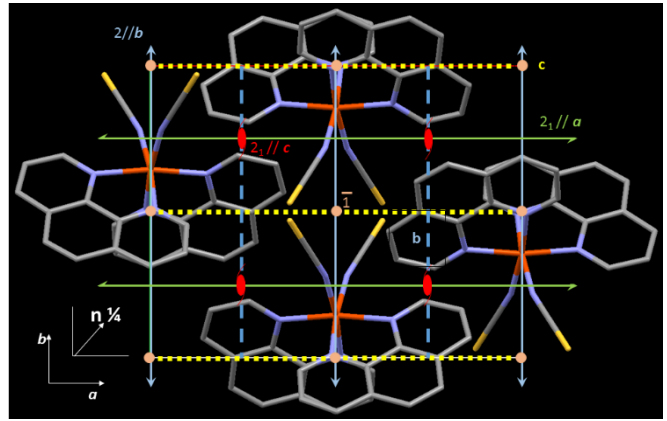

**Figure 3.** Crystal structure of the [Fe(phen)$_2$(NCS)$_2$] SCO material. For clarity, only few molecules are shown for clarity. The molecules lie on an axis 2 // **b** and are equivalent to each other by 2$_1$ screw axes, the b, c and n glide planes and inversion symmetry ($\bar{1}$).

## Crystalline point group

| $D_{2h}$ (mmm) | E | $C_2(z)$ | $C_2(y)$ | $C_2(x)$ | i | $\sigma(xy)$ | $\sigma(xz)$ | $\sigma(yz)$ | | |
|---|---|---|---|---|---|---|---|---|---|---|
| $A_g$ | 1 | 1 | 1 | 1 | 1 | 1 | 1 | 1 | | $x^2, y^2, z^2$ |
| $B_{1g}$ | 1 | 1 | −1 | −1 | 1 | 1 | −1 | −1 | $R_z$ | xy |
| $B_{2g}$ | 1 | −1 | 1 | −1 | 1 | −1 | 1 | −1 | $R_y$ | xz |
| $B_{3g}$ | 1 | −1 | −1 | 1 | 1 | −1 | −1 | 1 | $R_x$ | yz |
| $A_u$ | 1 | 1 | 1 | 1 | −1 | −1 | −1 | −1 | | |
| $B_{1u}$ | 1 | 1 | −1 | −1 | −1 | −1 | 1 | 1 | z | |
| $B_{2u}$ | 1 | −1 | 1 | −1 | −1 | 1 | −1 | 1 | y | |
| $B_{3u}$ | 1 | −1 | −1 | 1 | −1 | 1 | 1 | −1 | x | |

**Figure 4.** Character table of the *mmm* point group of the crystalline structure of [Fe(phen)$_2$(NCS)$_2$] crystal. We highlight the B$_{2u}$ (red) and B$_{3u}$ (blue) symmetry modes, selectively measured by polarized TD-THZ spectroscopy.

This is the case of the molecular breathing mode (Fig. 2), for which a totally symmetric Fe−N stretching occurs (in phase for the 6 Fe−N bonds), with almost rigid *phen* ligands [3,4]. Bending modes may also belong to the irreducible representation A. Vibrations (including other stretching and bending) belonging to the irreducible representation B are symmetric with respect to the identity operation *E*, but antisymmetric with respect to rotation around the $C_2$ axis. In addition, since the *phen* ligand is rigid, Fe−N elongation often couples to bending of the N-Fe-N groups and many modes have both stretching and bending characters [4]. Figure 2 shows very schematic representations of molecular motions for different symmetries.

The Fermi's golden rule states that a vibrational mode in a sample is "IR active", when it is associated with changes in the dipole moment, because of the coupling of this dipole moment change with the electric field of the electromagnetic radiation. For symmetry consideration, the coupling is allowed when the oscillatory dipole moment has the same symmetry as a component of the electric dipole vector (*x, y, z*). For the [Fe(phen)$_2$(NCS)$_2$] molecule, these modes may therefore belong to the irreducible representation A, when the resulting dipole moment created by the activation of the mode oscillates parallel to **b** (i.e. the $C_2$ molecular axis with a *y* component), or to the irreducible representation B (*x,z* component). Molecular vibrations are Raman active when the polarizability tensor transforms similar to quadratic functions of *x, y,* and *z*. As indicated in Fig. 1, for the isolated [Fe(phen)$_2$(NCS)$_2$] molecule of $C_2$ point group, molecular vibration modes of A and B symmetry are both Raman and IR active.

### 3.2 Symmetry of the crystalline modes

The [Fe(phen)$_2$(NCS)$_2$] SCO material crystallises in the *Pbcn* space group. Figure 3 shows the molecular packing and the symmetry operators within the unit cell. In the crystal, the $C_2$ molecular symmetry is kept, as the [Fe(phen)$_2$(NCS)$_2$] molecules lie on the Wyckoff position of $C_2$ symmetry, with the 2 fold axis corresponding to the crystalline axis **b**. The symmetry operators of the space group include also the screw axis 2$_1$//**a** and 2$_1$//**c**, the glide planes b ⊥ **a**, c ⊥ **b** and n ⊥ **c** as well as inversion symmetry in between molecules. All the molecules within the unit cell are symmetry equivalent and the asymmetric unit is made of half a [Fe(phen)$_2$(NCS)$_2$] molecule.

The point group of the *Pbcn* space group is *mmm* (D$_{2h}$) and the corresponding character table is shown in Fig. 4. Low frequency phonon modes in the crystal can be discussed in terms of molecular stretching and/or bending. The symmetry of the modes is related to relative atomic motions within a molecule around its $C_2$ axis or from molecule to molecule, preserving or not the symmetry operators of the space group. The DFT calculations provide then for each mode three important characteristics:
- the frequency of the modes, which can be compared to the ones measured experimentally
- the nature of the modes in term of atomic motions (stretching, bending, torsion…)
- the symmetry of the modes through the eigenvector.

The visualization of the relative motions of atoms (supplementary videos), and their equivalence by symmetry operators of the space group, allows for identifying the (pseudo) symmetry of the calculated modes. As inversion symmetry operator lies in between two molecules (Figure 3), their in-phase or out-of-phase distortions with respect to inversion symmetry allows distinguishing easily "*u*" and "*g*" modes. Figure 5 shows how to attribute the symmetry of the modes by observing the relative motions of atoms and comparing with the character table (Fig. 4). A$_g$ modes preserve all the symmetry operators of the crystalline structure, with in-phase motions of all the molecules in the crystal. B$_{2u}$ modes (Fig. 4) preserve the $C_2(y)$ molecular symmetry axis, as well as the b ⊥ **a** and n ⊥ **c** glide planes. B$_{3u}$ modes preserve $C_2(x)$ symmetry axis of the *mmm* point group (2$_1$ screw axis // **a** in the crystal), as well as the c ⊥ **b** and n ⊥ **c** glide planes. In a similar way: A$_u$ modes respect the $C_2(y)$ and 2$_1$ axis, but not the b, c nor n glide planes. B$_{1g}$ modes preserve the n glide plane ($\sigma(xy)$), B$_{2g}$ modes preserve the c glide plane ($\sigma(xz)$), B$_{3g}$ modes preserve the b glide plane ($\sigma(yz)$) and so on.

# Phonon modes of the spin crossover complex [Fe(phen)$_2$(NCS)$_2$]

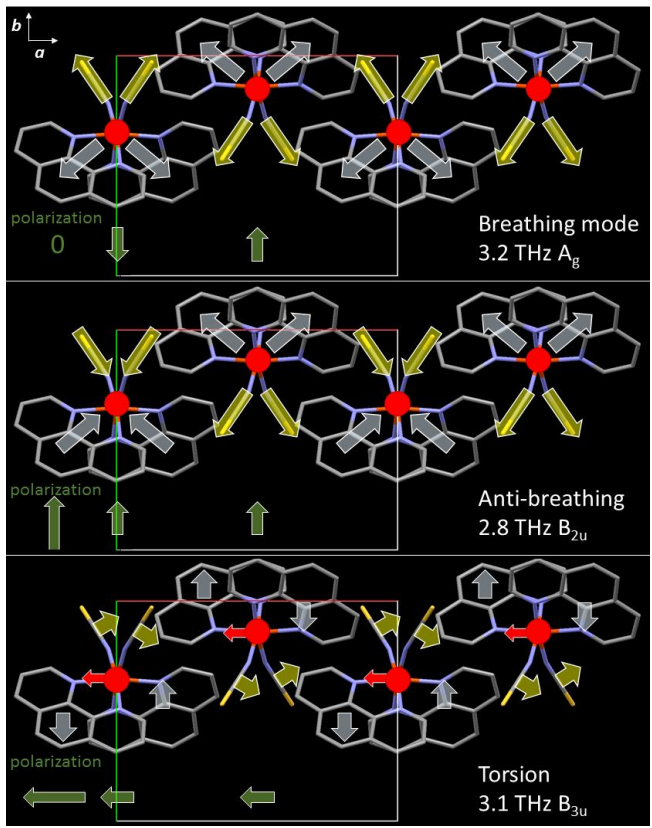

**Figure 5.** Lattice phonon modes of the [Fe(phen)$_2$(NCS)$_2$] crystal of $A_g$, $B_{2u}$ and $B_{3u}$ symmetries. Green arrows show polarization change for each molecule and the total polarization change within the unit cell (left). Only few molecules are shown for clarity.

The character table in Fig. 4 indicates 3 types of IR-active modes $B_{3u}$, $B_{2u}$ and $B_{1u}$, with polarization along *x*, *y*, or *z* respectively, and a Raman-active modes of symmetry $A_g$ transforming like $x^2$, $y^2$, $z^2$ and 3 Raman modes $B_{3g}$, $B_{2g}$ and $B_{1g}$ and transforming like *xy*, *xz* or *yz*. With these symmetry considerations in mind, we discuss now IR, THz, and Raman spectroscopy data, and try to assign the proper calculated vibrational modes to experimental observations.

### 3.3 N–CS stretching bands of [Fe(phen)$_2$(NCS)$_2$] crystal

In previous works on crystals of [Fe(phen)$_2$(NCS)$_2$], the bands observed round ≈ 2064 – 2076 cm$^{-1}$ by IR and ≈ 2072 cm$^{-1}$ by Raman were attributed to N–CS stretching modes, as supported by comparison with DFT calculation on isolated [Fe(phen)$_2$(NCS)$_2$] molecule [5,13,22,40].

For the isolated molecule of $C_2$ symmetry, there are 2 N–CS stretching modes:
- the symmetric A mode calculated around 2040 – 2112 cm$^{-1}$
- the antisymmetric B mode calculated around 2028 – 2096 cm$^{-1}$.

The unit cell in the crystal contains 4 molecules and there are 8 N–CS stretching modes of symmetry $A_g$, $A_u$, $B_{1g}$, $B_{2g}$, $B_{3g}$, $B_{1u}$, $B_{2u}$ or $B_{3u}$.

Raman and IR measurements shown in Fig. 6 are consistent with previous results reporting well-identified and very intense N–CS modes, often used has marker of the spin state [5]. One can distinguish 3 IR bands at 2056, 2061 and 2073 cm$^{-1}$ in the experimental data. The Raman band around 2070 cm$^{-1}$ has asymmetric shape and the fit involves 4 modes at 2059, 2065, 2071 and 2076 cm$^{-1}$. The DFT calculations in the solid state allow to describe the eigenvectors and the symmetries of the 8 N–CS stretching phonon modes (Table 1 and supplementary videos) found around 2000 cm$^{-1}$, *i.e.* slightly below experimental frequencies. These modes involve only the motions of the N and C atoms of the NCS group. We experimentally identify in Table 1 the 4 "*g*" Raman active and the 3 "*u*" IR active modes. As these measurements have no polarization analysis, we cannot identify their symmetries. We therefore tentatively assign the symmetry of the modes by sorting them in frequency order, for the observed and calculated "*u*" and "*g*" phonons. The overall ≈20 cm$^{-1}$ splitting of the N–CS stretching modes observed by IR and Raman is consistent with calculations. The $A_u$ mode is not active.

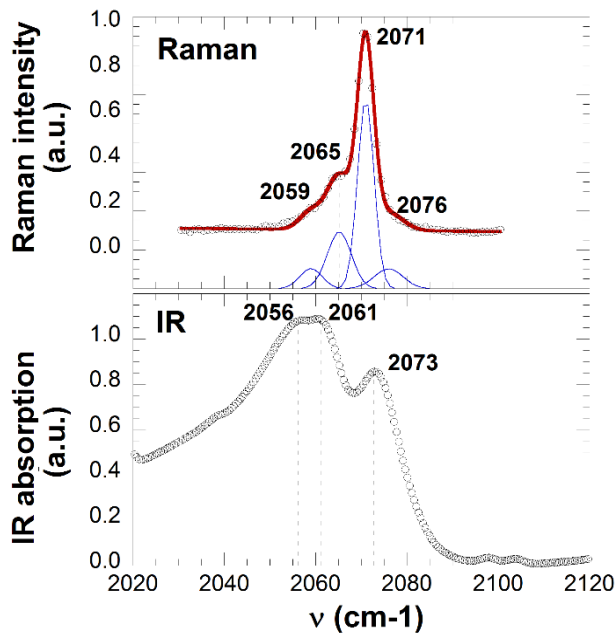

**Figure 6.** Measured IR and Raman N–CS stretching modes of the HS state.

**Table 1.** N–CS stretching mode in the HS state for [Fe(phen)$_2$(NCS)$_2$] calculated in the solid state and measured in Fig. 6. Videos of the modes are provided in supplemental material.

| Mode N° | ν(THz) calculated | ν(cm$^{-1}$) calculated | ν(cm$^{-1}$) measured | Symmetry and activity |
|---|---|---|---|---|
| 413 | 59.33 | 1979 | 2056 | $B_{3u}$ IR |
| 414 | 59.35 | 1980 | 2061 | $B_{1u}$ IR |
| 415 | 59.48 | 1984 | 2059 | $B_{3g}$ Raman |
| 416 | 59.49 | 1984 | 2065 | $B_{1g}$ Raman |
| 417 | 59.83 | 1996 | – | $A_u$ – |
| 418 | 59.89 | 1998 | 2073 | $B_{2u}$ IR |
| 419 | 60.00 | 2001 | 2071 | $B_{2g}$ Raman |
| 420 | 60.04 | 2003 | 2076 | $A_g$ Raman |

### 3.4 Low frequency modes and breathing modes

The low frequency modes in the THz range are known to have an important contribution to phase transition at thermal equilibrium, as their contribution to vibrational entropy is high. They also play a central role in the ultrafast structural dynamics involved in the LIESST phenomena. Indeed, the activation and damping of the molecular breathing mode and its coupling to torsion and lower frequency modes is the key feature of this process explaining both efficiency and speed [25,26,28,41]. Such coherent vibration during the photo-switching of [Fe(phen)$_2$(NCS)$_2$] were observed by femtosecond optical spectroscopy [4].

# Phonon modes of the spin crossover complex [Fe(phen)$_2$(NCS)$_2$]

Depending on the probe polarization (parallel or perpendicular to the molecular axis) the breathing ($\approx$110 cm$^{-1}$) and the bending ($\approx$85 cm$^{-1}$) modes can be detected during LIESST, as the photoexcited LS molecule reaches the HS state. The breathing mode, related with the elongation of the average Fe-N bond length, is activated first and energy is then transferred to the torsion of the N-Fe-N angles [42].Such ultrafast activation and damping of the molecular breathing mode during LIESST was observed in different systems by ultrafast optical and X-ray spectroscopies [4,25,26,28,41].

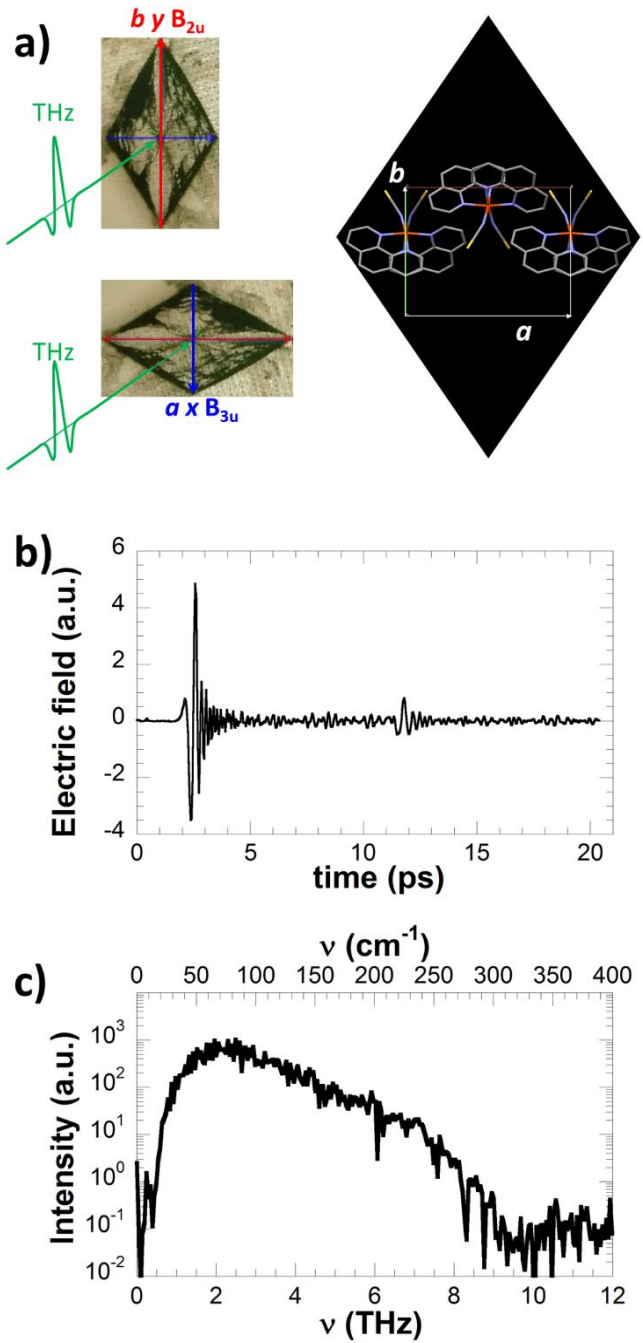

**Figure 7.** a) Schematic set-up for THz transmission, measured on lozenge-shape single crystal of [Fe(phen)$_2$(NCS)$_2$]. The thickness corresponds to the *c* axis and THz polarization was set along *b* (B$_{2u}$ modes) or *a* (B$_{3u}$). b) Time dependence of the electric field of the THz radiation c) Spectrum of the THz radiation obtained by FFT.

For better understanding these low frequency HS modes in crystal, we used Raman and TD-THz spectroscopy measurements. We took advantage of the linear polarization of the THz radiation for performing measurements on single crystals of [Fe(phen)$_2$(NCS)$_2$], allowing to select the symmetry of the measured modes. Crystals form plates with a characteristic lozenge shape (Fig. 7). The direction perpendicular to the surface corresponds to the crystalline axis *c* (thickness $\approx$ 30 µm). The short lozenge axis ($\approx$ 500 µm) corresponds to the crystalline axis *a*, and the long axis ($\approx$ 900 µm) corresponds to the crystalline axis *b* [3,4]. The THz radiation spot was focussed down to $\approx$ 300 µm diameter on the crystal near diffraction limit. Fig. 7b shows the time dependence of the electric field of the THz radiation and Fig. 7c shows its corresponding FFT spectrum, which allows probing "IR active modes" in the 0.5–8 THz range ($\approx$ 20–270 cm$^{-1}$). We performed TD-THz spectroscopy on single crystal, with THz electric polarization along the *a* crystalline axis for probing the B$_{3u}$ modes and with THz polarization along the *b* crystalline axis for probing the B$_{2u}$ modes. The THz absorption spectra shown in Fig. 8a differs for both polarizations. For the configuration probing B$_{2u}$ symmetry modes, we observed an intense mode at 93 cm$^{-1}$ and weaker ones around 57, 80, 107, 133, 176, 207 and 233 cm$^{-1}$. When probing B$_{3u}$ symmetry phonons, we observe an intense mode around 103 cm$^{-1}$ and weaker ones around 67, 170, 220, 235 and 257 cm$^{-1}$.

The low-frequency Raman measurements shown in Fig. 8b, reveals numerous Raman lines. Because of experimental limitation, detailed polarization analysis could not be performed, but we noticed that the relative intensity of several modes changes when the laser polarization is parallel to the *b* or to the *c* crystalline axis, as expected for an orthorhombic system. Hereafter we assign to each low frequency mode, observed by TD-THz data for B$_{2u}$ and B$_{3u}$ modes and by Raman, a mode found by DFT calculation (Table 2), relevant in term of frequency, symmetry and nature of the atomic motions. We provide videos for different modes in supplementary materials. The differentiation between "*g*" Raman-active and "*u*" IR-active modes is easy for comparing calculations and measurements. Hereafter the discussion is restricted to only few modes.

The totally symmetric breathing mode (A$_g$) is a Raman active mode, associated with the in-phase breathing of all the molecules in the crystal (Fig. 5). In DFT calculation, it corresponds to the mode n°88 (frequency calculated at $\nu_c$=3.19 THz / 106 cm$^{-1}$, see supplementary video) and matches well with the Raman mode measured at 105 cm$^{-1}$. The amplitude of this mode is strong when the laser polarization is // *b* and weak when // *c*. We attribute this feature to a resonant effect, as the 633 nm laser excitation is close to the $t_{2g}$-$e_g$ electronic gap in this system. This electronic transition is strong when light polarization is parallel to the molecular axis *b* and is strongly coupled to the breathing mode as underlined by time-resolved spectroscopy and TD-DFT calculation [3]. For this totally symmetric breathing mode, the intra-molecular motions are invariant with respect to the $C_2$ molecular axis, which correspond to the A symmetry breathing mode of the molecular point group. In the crystal, the in-phase breathing of the molecules corresponds to the totally symmetric A$_g$ breathing mode.

For the B$_{2u}$ mode n° 80 ($\nu_c$=2.82 THz / 94 cm$^{-1}$), the intra-molecular motions are also invariant with respect to the $C_2$ molecular axis and correspond also to the A symmetry breathing of the molecular point group. However, the oscillations are out-of-phase between molecules related by inversion symmetry.

# Phonon modes of the spin crossover complex [Fe(phen)$_2$(NCS)$_2$]

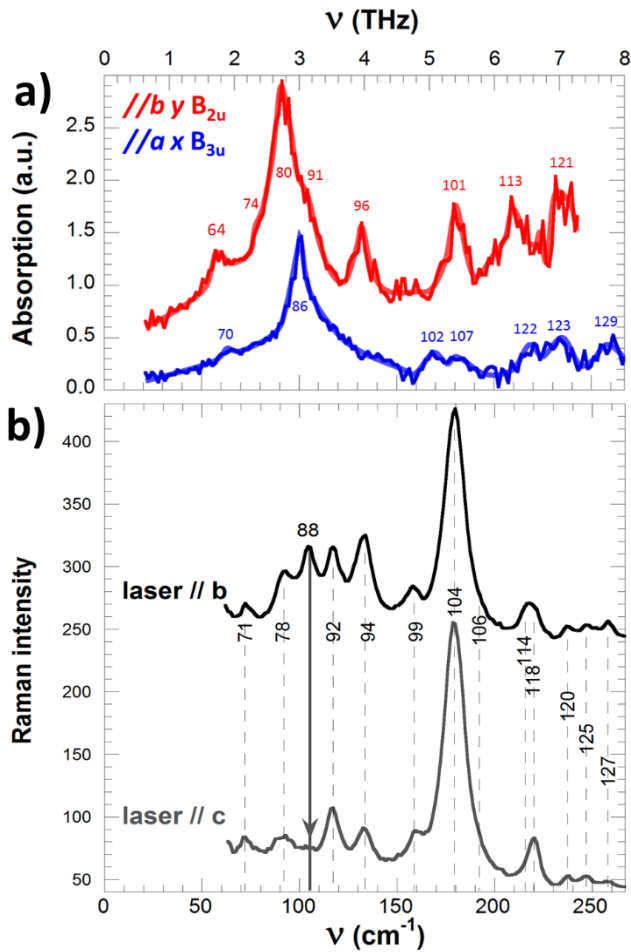

**Figure 8.** a) THz absorption measured in the 20-270 cm$^{-1}$ range on a single crystal of [Fe(phen)$_2$(NCS)$_2$] for THz polarization along ***a*** and ***b*** probing B$_{3u}$ and B$_{2u}$ modes. b) Measured Raman spectra with laser excitation // ***b*** or // ***c***. Each band is tentatively assigned by a n° referring to a mode calculated by DFT with appropriate symmetry and frequency (see Table 2 for more details).

This antisymmetric breathing B$_{2u}$ mode corresponds to the intense mode observed at 2.8 THz (93 cm$^{-1}$) and shown in Fig. 5 and supplementary video. This is the only B$_{2u}$ mode calculated in this frequency range and it his mode has a strong THz activity. Indeed, the molecular dipoles oriented along the molecular axis ***b*** and are strongly modulated by the large amplitude molecular breathing mode, with the in-phase Fe–phen and Fe–NCS elongations. It should be noticed that due to the neutral character of the molecule and the oxidation state of iron is (II), electrons are transferred to the NCS groups and the bonds have strong ionic character Fe$^{2+}$–(NCS$^-$)$_2$ [4]. By symmetry, the dipole moment *d* of the molecule is oriented along ***b*** and our calculations with Gaussian 09 give *d* = 18 D in the HS state. Due to the out-of-phase breathing of the molecules head-to-tail along the ***b*** axis, the induced variations of their dipole moments contribute to a total change along the *y* component (Fig. 5). The molecular breathing nature of the A$_g$ mode n°88 and B$_{2u}$ mode n°80 are therefore similar, but the frequency of the A$_g$ breathing mode is higher than the one of the antisymmetric B$_{2u}$ breathing mode. It can be explained by the fact that when molecules breathe out-of-phase (B$_{2u}$), the total volume occupied within the unit cell remains constant and intermolecular contact remains similar. When molecules expand in phase (A$_g$), the distance between molecules decreases, hence making interactions stronger. The potential energy surface along the antisymmetric breathing B$_{2u}$ mode (≈93 cm$^{-1}$) is therefore softer than along the A$_g$ breathing mode (≈106 cm$^{-1}$). We can underline that the A$_g$ breathing mode has calculated and observed frequencies slightly lower than the one of the breathing mode observed during LIESST. Indeed, ultrafast studies of LIESST are performed in a LS phase: when molecules locally switch to HS state within less than 1 ps, the unit cell still holds LS parameters as it has no time to expand toward the HS cell parameters on this timescale [2,29]. The molecular potential energy curve of the photoinduced HS molecule in the LS lattice is harder (110 cm$^{-1}$) than the one of the HS molecule in an expended HS lattice at thermal equilibrium.

In addition to these modes, many other modes exist and correspond to a mixture of molecular breathing and bending, like the mode B$_{2g}$ n° 104 ($\nu_c$=5.24 THz / 175 cm$^{-1}$), for which the molecules breathing and bending out-of-phase are the ones related by the ***b*** glide plane. Its intra-molecular dynamics is equivalent to the one of the B$_{2u}$ mode n° 101 ($\nu_c$=5.19 THz / 176 cm$^{-1}$), for which the molecules breathing and bending out-of-phase are the ones linked by the inversion symmetry.

The intense B$_{3u}$ mode at 3.1 THz (103 cm$^{-1}$) is attributed to the torsion mode n° 86 ($\nu_c$=3.1 THz / 103 cm$^{-1}$) represented in Fig. 5 and shown in supplementary video as well. The Fe$^{II}$ ions move in phase along the ***a*** axis, whereas the NCS$^-$ fragments bend in opposite direction compare to the motion of the Fe$^{2+}$ cations. This mode therefore also strongly modulates the component of the polarization along *x*, explaining why it is so intense (Fig. 8b).

Low frequency modes are listed in Table 2, where we assign the calculated modes to the ones measured and described in terms of stretching, bending or torsion.

## 4 Conclusions

This combined experimental – theoretical study illustrates that some lattice phonon modes in molecular crystals made of several symmetry-equivalent molecules in the unit cell, can be viewed as a combination of identical intra-molecular vibrations, in phase or out-of-phase from molecules to molecules. These combinations, with respect to the different symmetry operators of the space group, correspond to various modes of different symmetries. Compared to the simple view used to describe vibrations in SCO materials from a single molecular perspective, the present study illustrates how tacking into account molecular packing and symmetry allows for a complete description of the phonon modes. This is true for high frequency modes like the N–CS stretching phonons, for which 8 lattice modes exist and 7 are spectroscopically active. It also stands for some lower frequency modes like breathing or bending modes, which may be symmetric or antisymmetric. The duplication of modes of different symmetry in the crystal, compared to the isolated molecule, may play an important role for the description of the vibrational entropy contribution to phase transition at thermal equilibrium. The results also show that polarization-resolved TD-THz spectroscopy performed on single crystals allows for a clear identification of the symmetry of the phonon modes. More interestingly, some of these modes are known to be strongly coupled to the change of spin state. With the development of intense and pulsed THz and IR lasers, allowing for non-linear phononics,[30,31] our goal now is to excite selectively these mode for driving spin-state switching.

# Phonon modes of the spin crossover complex [Fe(phen)$_2$(NCS)$_2$]

**Table 2.** Low frequency modes, IR or Raman active, sorted by increasing frequency of the calculated modes ($\nu_c$), and compared with measured frequencies ($\nu_m$). The symmetry of the modes and the nature of the vibrations are also indicated. "Symmetric" and "asymmetric" refer to the molecular C$_2$ axis, whereas anti-symmetric is used for out-of phase vibration of molecules with respect to inversion symmetry between them. Supplemental videos show the modes marked by and a star

| Active | Mode N° | $\nu_c$ (cm$^{-1}$/THz) | $\nu_m$ (cm$^{-1}$/THz) | Symmetry | Nature of the mode |
|---|---|---|---|---|---|
| IR | 64 | 58 / 1.74 | 57 / 1.7 | y, B$_{2u}$ | Symmetric torsion of the FeN$_6$ octahedron |
| IR | 70 | 68 / 2.05 | 67 / 2.0 | z, B3$_u$ | Asymmetric torsion of the octahedron |
| Raman | 71 | 73 / 2.20 | 72 / 2.16 | A$_g$ | Symmetric FeN$_6$ torsion and bending of the Fe-NCS bonds |
| IR | 74 | 80 / 2.40 | 80 / 2.4 | y, B$_{2u}$ | Symmetric FeN$_6$ torsion and bending of the Fe-NCS bonds |
| Raman | 78 | 88 / 2.64 | 92 / 2.76 | xy, B$_{1g}$ | Libration of the FeN$_6$ octahedron around Fe atom |
| IR | 80* | 94 / 2.82 | 93 / 2.80 | y, B$_{2u}$ | Antisymmetric breathing mode with symmetric FeN$_6$ breathing |
| IR | 86* | 103 / 3.10 | 103 / 3.1 | x, B$_{3u}$ | Asymmetric Fe-N$_{phen}$ stretching, and Fe-NCS torsion |
| Raman | 88* | 106 / 3.19 | 106 / 3.18 | A$_g$ | Symmetric Breathing mode |
| IR | 91* | 108 / 3.26 | 107 / 3.2 | y, B$_{2u}$ | Butterfly mode: symmetric Phen bending & Fe-NCS torsion |
| Raman | 92 | 108 / 3.26 | 117 / 3.51 | yz, B$_{3g}$ | Asymmetric Fe-N$_{phen}$ stretching, and Fe-NCS torsion |
| Raman | 94 | 145 / 4.36 | 133 / 3.99 | yz, B$_{3g}$ | Fe-NCS torsion |
| IR | 96 | 147 / 4.40 | 133 / 4.00 | y, B$_{2u}$ | Symmetric Fe-NCS torsion |
| Raman | 99 | 150 / 4.49 | 158 / 4.74 | xz, B$_{1g}$ | Asymmetric Fe-NCS torsion |
| IR | 101* | 173 / 5.19 | 176 / 5.40 | y, B$_{2u}$ | Symmetric stretching of Fe-N$_{phen}$, Fe-NCS, CS & N-Fe-N bending |
| IR | 102 | 174 / 5.21 | 170 / 5.10 | x, B$_{3u}$ | Asymmetric stretching of Fe-N$_{phen}$, Fe-NCS, CS & N-Fe-N bending |
| Raman | 104* | 175 / 5.24 | 175 / 5.25 | xz, B$_{2g}$ | Symmetric stretching of Fe-N$_{phen}$, Fe-NCS, CS & N-Fe-N bending |
| Raman | 106 | 176 / 5.28 | 184 / 5.51 | yz, B$_{3g}$ | Symmetric stretching of Fe-N$_{phen}$, Fe-NCS, CS & N-Fe-N bending |
| IR | 107 | 177 / 5.30 | 170 / 5.10 | x, B$_{3u}$ | Asymmetric Fe-NCS stretching |
| IR | 113 | 207 / 6.22 | 207 / 6.20 | y, B$_{2u}$ | Symmetric Fe-N$_{phen}$ stretching, Fe-NCS and ligand |
| Raman | 114 | 208 / 6.25 | 217 / 6.50 | xz, B$_{2g}$ | Symmetric Fe-N$_{phen}$ stretching, Fe-NCS and ligand bending |
| Raman | 118 | 225 / 6.74 | 222 / 6.65 | xz, B$_{2g}$ | Symmetric Fe-N$_{phen}$ stretching, Fe-NCS and ligand bending |
| Raman | 120 | 225 / 6.75 | 237 / 7.10 | A$_g$ | Symmetric Fe-N$_{phen}$ stretching, Fe-NCS and ligand bending |
| IR | 121 | 226 / 6.78 | 233 / 7.00 | y, B$_{2u}$ | Antisymmetric global stretching mode of the molecule |
| IR | 122 | 226 / 6.79 | 220 / 6.8 | x, B$_{3u}$ | Asymmetric Fe-NCS stretching FeN$_6$ libration |
| IR | 123 | 226 / 6.79 | 235 / 7.05 | x, B$_{3u}$ | Asymmetric Fe-NCS stretching FeN$_6$ libration |
| Raman | 125 | 258 / 7.73 | 248 / 7.43 | xz, B$_{2g}$ | FeN$_6$ bending, CS stretching and ligand butterfly mode |
| Raman | 127 | 260 / 7.8 | 258 / 7.73 | A$_g$ | FeN$_6$ bending, CS stretching and ligand butterfly mode |

# Phonon modes of the spin crossover complex [Fe(phen)$_2$(NCS)$_2$]

## Acknowledgements

Parts of this research were performed within the international laboratory LIA IM-LED (CNRS). We thank the SIR platform of ScanMAT at Univ. Rennes 1 for technical assistance for the Raman measurements and the ANR (ANR-13-BS04-0002 FEMTOMAT) for financial support.

## Author contribution statement

E. Collet conceived the project. G. Azzolina, T. Ichii, L. Guerin, E. Collet and K. Tanaka performed the THz measurements. J. Kubicki performed IR measurements. A. Moréac performed Raman measurements. N. Daro and G. Chastanet synthesized the single crystals. S.F. Matar performed density functional theory calculations. E. Collet, R. Bertoni and M. Cammarata investigated the phonon modes. E. Collet wrote the paper and all authors contributed to discussions and gave comments on the manuscript.